\documentclass[a4paper,12pt]{article}
\usepackage{graphicx,epsf}

\begin{document}
\title{Nonlinear perturbations for dissipative
and interacting relativistic fluids}
\author{David Langlois$^{1,2}$, Filippo Vernizzi$^3$\\
{\small {}}\\
{\small ${}^1${\it APC (Astroparticules et Cosmologie),}}\\
{\small {\it
UMR 7164 (CNRS, Universit\'e Paris 7, CEA, Observatoire de
Paris)}}\\
{\small {\it  11 Place Marcelin Berthelot, F-75005 Paris, France;}}\\
{\small ${}^2${\it GReCO, Institut d'Astrophysique de Paris, CNRS,}}\\
{\small {\it 98bis Boulevard Arago, 75014 Paris, France;  }}\\
{\small {\it and}}\\
{\small ${}^3${\it Helsinki Institute of Physics, P.O. Box 64,}}\\
{\small {\it FIN-00014 University of Helsinki - Finland}}\\
}

\date{\today}
\maketitle

\def\beq{\begin{equation}}
\def\eeq{\end{equation}}
\newcommand{\bea}{\begin{eqnarray}}
\newcommand{\eea}{\end{eqnarray}}
\def\bi{\begin{itemize}}
\def\ei{\end{itemize}}
\def\Tdot#1{{{#1}^{\hbox{.}}}}
\def\Tddot#1{{{#1}^{\hbox{..}}}}
\def\D{{\cal D}}
\def\d{{\delta}}
\def\T{{\bf T}}
\def\perp{n}
\def\R{{\cal K}}
\def\L{{\cal L}_u}
\def\HH{{\cal H}}
\def\p{p}
\def\q{q}

\begin{abstract}
We develop a covariant formalism to  study nonlinear perturbations
of dissipative and interacting relativistic fluids. We derive
nonlinear evolution equations for various covectors defined as
linear combinations of the spatial gradients of the local number
of e-folds and of some scalar quantities characterizing the fluid,
 such  as the energy density or the particle number density. For
interacting fluids we decompose perturbations into adiabatic and
entropy components and derive their coupled evolution equations,
recovering and extending the results obtained in the
context of the linear theory. For
non-dissipative and noninteracting fluids, these evolution
equations reduce to the conservation equations that we have
obtained in recent works.
We also illustrate geometrically
the meaning of the covectors that we have introduced.
\end{abstract}

\newpage

\section{Introduction}
Recently, we have developed \cite{lv05a,lv05b} a new  formalism
 for nonlinear relativistic  perturbations, based on a
covariant approach \cite{Hawking:1966qi,ellis}  and inspired by
the work of Ellis and Bruni \cite{Ellis:1989jt}.  A key quantity
in our formalism is the covector $\zeta_a$, a linear combination
of the spatial gradients of the energy density $\rho$ and of
$\alpha$, the local number of e-folds, or integrated expansion
along each worldline of the fluid.

As we showed in \cite{lv05a,lv05b}, $\zeta_a$, for a barotropic
fluid, obeys a remarkably simple {\it conservation equation},
which is {\it exact and fully nonlinear}. In the linear
approximation, this conservation equation reduces  to the usual
conservation law for the linear curvature perturbation on uniform
energy density hypersurfaces, usually denoted $\zeta$. Thus, our
covector $\zeta_a$ can be seen as a covariant and nonlinear
generalization of the usual $\zeta$.

It must be stressed that, although our initial motivations were
related to  cosmology, our formalism applies, in fact, to any
relativistic fluid, whatever the underlying geometry. Thus, in the
following we will not assume any particular type of spacetime,
unless specified.

 The purpose of the present work is to clarify the geometrical
meaning of $\zeta_a$, and to extend our formalism in two new
directions. Whereas our formalism has been developed so far for a
{\it perfect} fluid, we consider here its extension to the case of
dissipative relativistic fluids.
Moreover, we consider the possibility of having
several different {\em interacting} fluids.

The covariant formalism of Ellis and Bruni \cite{Ellis:1989jt}
was first extended
to a dissipative multifluid system
in  \cite{Hwang:1990am,Dunsby:1991vv}, where
both the exact and the linear form of perturbation equations were given
for noninteracting fluids. It was later generalized to the case of interacting
fluids in \cite{Dunsby:1991xk} for {\em linear} perturbations.

The extension to interacting fluids is particularly
useful in cosmology, where several types of matter coexist. As
a typical example, one can mention the analysis of the
cosmic microwave background
anisotropies, which has been for instance considered
 within the covariant formalism and {\em at linear order} in
\cite{Challinor}.

In this paper we work in the framework of our nonlinear formalism
\cite{lv05a,lv05b}, based on covectors such as $\zeta_a$. As in
the single perfect fluid case, we show that it is possible to
derive simple, exact, and fully nonlinear evolution equations for
various covectors, constructed as linear combinations of the
spatial gradients of scalar quantities that characterize the
fluids, such as the energy density or the number particle, and of
the local number of e-folds $\alpha$.

The  equations we obtain
represent nonlinear generalizations
of linear perturbation equations, some of which  have been already
studied in the context of linear cosmological perturbations (see,
in particular,
\cite{Malik:2002jb,Malik:2004tf}). However, our equations are
covariant, i.e., independent of the coordinate system, and nonlinear, and it
is straightforward
to linearize or expand them up to  second, or higher,
 order in the perturbations.
Furthermore, our covariant approach allows to identify easily which
properties belong specifically to the linear or second order expansion,
and which properties remain valid  at higher orders.

This work is organized as follows. In Sec.~\ref{sec:covariant} we
introduce the covariant formalism for dissipative fluids while in
Sec.~\ref{sec:covariantpert} we discuss the geometrical meaning of
the nonlinear covector $\zeta_a$. In Sec.~\ref{sec:identity} we
derive an identity that can be used to construct conservation
equations for various nonlinear covectors, representing the
perturbations of scalar quantities of a dissipative fluid. The
conservation equations for these quantities are derived in
Sec.~\ref{sec:nonconservation}, while in
Sec.~\ref{sec:interacting} we extend our formalism to interacting
fluids.  Finally, in Sec.~\ref{sec:conclusion}, we conclude.

\section{Covariant formalism}
\label{sec:covariant} In this section, we briefly review the
covariant description for a dissipative relativistic fluid. There
is a substantial  literature on dissipative relativistic fluids,
which is reviewed, for instance, in \cite{maartens}. The extension
of irreversible thermodynamics to relativistic fluids is hampered
by subtle issues. In particular, the first extensions, due to
Eckart in 1940 and to Landau and Lifshitz in the 1950's suffer
from non-causal behavior. An extended theory, which does not
suffer from this problem was developed by Israel and Stewart
\cite{Israel:1979wp}. In the present work, we will not need to
enter the details of these various formulations and, for details,
 we refer the reader to \cite{maartens}, whose presentation and notation
will be followed here.

We first define the unit four-velocity of the fluid $u^a$ as the
average velocity of the fluid particles. This means  that $u^a$ is
proportional to  the particle current $n^a$, which can thus be
written as
\beq
n^a=n u^a, \qquad u^a u_a=-1,
\eeq
where $n$ is the particle number density.

In addition to the particle number density $n$, the fluid is
characterized by local equilibrium scalars: the energy density
$\rho$, the pressure $\p$, the entropy $S$ and the temperature
$T$.  In
general the effective pressure deviates from the local equilibrium
pressure so that $\p_{\rm eff}= \p+\Pi$. The energy-momentum
tensor can be written in the form
\beq
T_{ab}=\rho u_a u_b+ \left(\p+\Pi\right) h_{ab}+ q_{a} u_{b}+q_b
u_a +\pi_{ab}, \label{emt}
\eeq
where $h_{ab}$ is the projection tensor orthogonal to the fluid velocity
$u^a$,
\beq
h_{ab}=g_{ab}+u_a u_b, \quad \quad (h^{a}_{\ b} h^b_{\ c}=h^a_{\
c}\, , \quad h_a^{\ b}u_b=0),
\eeq
and where the energy flow $q_a$ and the anisotropic stress
$\pi_{ab}$ satisfy the following properties:
\beq
q_au^a=0, \quad \pi_{ab}=\pi_{ba}, \quad \pi_a^{\ a}=0, \quad \pi_{ab}u^b=0.
\eeq

It is useful to introduce the familiar decomposition
\beq
\nabla_b u_a=\sigma_{ab}+\omega_{ab}+{1\over 3}\Theta
h_{ab}-\dot{u}_a u_b, \label{decomposition}
\eeq
with the (symmetric) shear tensor $\sigma_{ab}$, and the
(antisymmetric) vorticity  tensor $\omega_{ab}$; the volume
expansion, $\Theta$, is defined by
\beq
\Theta \equiv \nabla_a u^a,
\eeq
while $\dot u^a$ is the acceleration, with the dot denoting the
covariant derivative projected along $u^a$, i.e., $\dot{} \equiv
u^a \nabla_a $.

We also introduce the covariant spatial derivative, which is
defined as
\beq
\label{Da} D_a \chi \equiv h_a^{\ b}\nabla_b \chi, \quad \quad D_a
\chi_b=h_a^{\ c} h_b^{\ d} \nabla_c \chi_d, \quad \quad {\rm etc,}
\eeq
 where $\chi$ and $\chi_a$ are a generic scalar and covector, respectively.
As illustrated in \cite{Ellis:1989jt}, the covariant spatial
derivative is particularly useful to deal  with cosmological
perturbations in a {\sl covariant}  way, as an alternative to the
standard coordinate based approach.

The
conservation of the energy-momentum tensor,
\beq
\label{conserv} \nabla_a T^a_{\ b}=0,
\eeq
yields, by  projection along $u^a$, the energy conservation
equation,
\beq
\dot\rho + \Theta (\rho + \p+\Pi)+ \pi^b_{\ a} \nabla_b u^a +
\nabla_a q^a + q^a \dot u_a=0.
\eeq
Using the decomposition (\ref{decomposition}) and the definition
of the spatial covariant derivative (\ref{Da}), one  can rewrite
the above equation in the form
\beq
\dot\rho + \Theta (\rho + \p)= {\cal D}, \quad \quad {\cal
D}=-\left( \Theta\Pi+\pi^{ab} \sigma_{ab} + D_a q^a + 2 q^a \dot
u_a\right). \label{energy_cons}
\eeq
The scalar quantity ${\cal D}$ thus  contains all the dissipative
terms and vanishes   for a  perfect fluid.

In irreversible thermodynamics, the entropy is not conserved but
increases according to the second law of thermodynamics. This can
be expressed by the inequality
\beq
\nabla_a S^a\geq 0,
\eeq
where $S^a$ is the entropy current.
One usually writes $S^a$ in the form
\beq
S^a=Snu^a+\frac{R^a}{T},
\eeq
where $R^a$ is a dissipative term.  As discussed in \cite{maartens},
the explicit form for $R^a$ varies
according to the formalisms which have been introduced in the
literature. The entropy $S$
and temperature $T$ are the local equilibrium quantities, which
are related via the Gibbs equation,
\beq
TdS=d\left(\frac{\rho}{n}\right)+\p\, d\left(\frac{1}{n}\right).
\eeq
This implies, in particular,
\beq
nT\dot S=\dot\rho+\Theta \left(\rho+\p\right),
\eeq
where we have assumed conservation of the particle number, i.e.
\beq
\nabla_an^a=0 \quad \Leftrightarrow \quad \dot n+\Theta n=0.
\eeq

Using the energy conservation equation (\ref{energy_cons}),
this can be rewritten
as
\beq
nT\dot S=\D.
\eeq
In terms of the entropy density $s=nS$, this gives, using once more
the particle conservation equation,
\beq
\label{s_dot}
\dot s+\Theta s=\frac{\D}{T}.
\eeq
Using this equation,  one
gets
\beq
\nabla_a S^a=\frac{\D}{T}+\nabla_a\left(\frac{R^a}{T}\right).
\eeq
In the case of a non-dissipative fluid,
the right hand side is zero and the
above relation then expresses the conservation of entropy.

\section{Nonlinear covector}
\label{sec:covariantpert}

Here we illustrate the geometrical meaning of the nonlinear
covector that we introduced in \cite{lv05a,lv05b}. This
interpretation can easily be extended  to the other covectors of
the same form that will be defined in this paper.

As we showed in our recent works \cite{lv05a,lv05b} (see also
\cite{Lyth:2003im,Lyth:2004gb,Rigopoulos:2003ak}
for other recent formulations of
conserved nonlinear perturbations), a crucial
quantity to define conserved nonlinear perturbations is the
spatial gradient of the local number of e-folds, $\alpha$, which
is defined as the integration of $\Theta$ along the fluid world
lines with respect to the proper time $\tau$,
\beq
\label{alpha_def} \alpha \equiv {1\over 3}\int d\tau \, \Theta.
\eeq
It follows that
\beq
\Theta = 3 \dot\alpha = 3 u^a\nabla_a\alpha  .
\eeq

 In \cite{lv05a,lv05b}, we have introduced, for a {\em perfect} fluid,
a linear combination
of the spatial covariant derivatives of $\alpha$ and $\rho$,
\beq
\zeta_a= D_a \alpha-\frac{\dot\alpha}{\dot \rho} D_a \rho.
\label{zeta2} \label{zeta}
\eeq
This covector  is fully  conserved on all scales for
adiabatic perturbations, and  can be seen as the nonlinear
 generalization  of the usual $\zeta$.
In Sec.~\ref{sec:nonconservation} we will rederive the conservation
equation  for this quantity.

At this stage, it is instructive to give a graphical
representation of $\zeta_a$. To work with a scalar quantity rather
than a covector, let us consider an infinitesimal vector $e^a$
which is orthogonal to the fluid four-velocity $u^a$ at some
spacetime point $\q$. One can then write
\beq
e^a \zeta_a=\Delta\alpha-\frac{\dot\alpha}{\dot \rho}\Delta\rho,
\eeq
with
\beq
\Delta\alpha\equiv e^a\nabla_a \alpha, \quad \quad
\Delta\rho\equiv e^a\nabla_a \rho.
\eeq

\begin{figure}
\begin{center}
\includegraphics[height=15pc]{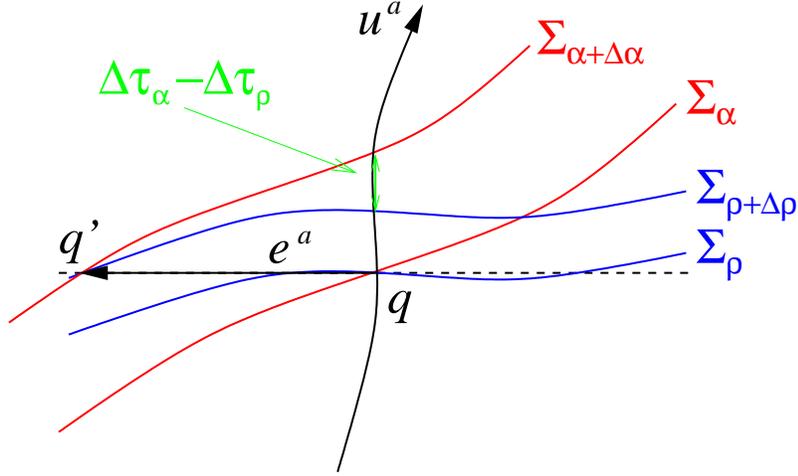}
\caption{Geometric interpretation of $\zeta_a$.} \label{fig}
\end{center}
\end{figure}

 As shown in Fig.~\ref{fig},
starting from our fiducial reference point $\q$, the infinitesimal
quantity $\Delta \alpha$ corresponds to the shift in $\alpha$ when
one goes from $\q$ to the neighboring point $\q'$ indicated by
$e^a$. Thus $\q'$ belongs to the hypersurface
$\Sigma_{\alpha+\Delta\alpha}$ characterized by the constant value
$\alpha+\Delta\alpha$. Similarly, $\q'$ belongs to the constant
energy density hypersurface $\Sigma_{\rho+\Delta\rho}$. Now, these
two hypersurfaces also intersect the fluid worldline that goes
through $\q$, but in general the intersections differ.
The quantity $e^a \zeta_a$ quantifies, in terms of the number of
e-folds, the separation between these two intersection points.
 Indeed, the
proper time interval between  $\q$ and the intersection of the
hypersurface $\Sigma_{\alpha+\Delta\alpha}$ with the worldline of $\q$
is $\Delta \tau_\alpha\equiv {\Delta \alpha}/{\dot\alpha}$ while the
proper time interval between  $\q$ and the intersection of the
hypersurface $\Sigma_{\rho+\Delta\rho}$ with the worldline of $\q$
is $\Delta \tau_\rho\equiv {\Delta \rho}/{\dot\rho}$. The difference
between these two proper time intervals is shown in the figure
and
the corresponding variation of $\alpha$ during this time difference interval
is $e^a \zeta_a= \dot \alpha (\Delta \tau_\alpha -
\Delta \tau_\rho)$.

\section{An identity for nonlinear covectors}
\label{sec:identity}

In this section we will derive a general identity  which we will
use later for various cases.  Namely, we will show that if one
starts with an equation of the form
\beq
\label{conserv_f}
\dot f+\Theta g=0,
\eeq
where $f$ and $g$ are two scalar quantities and, as before,  the
dot denotes the derivative along $u^a$, one finds the identity
\beq
\L \left(D_a \alpha- \frac{\dot\alpha}{\dot f}D_a f\right)= 3
\frac{\dot\alpha^2}{\dot f}\left( D_a g- \frac{\dot g}{\dot
f}D_af\right). \label{D_conserv_f3}
\eeq

To show this, let us start by rewriting Equation (\ref{conserv_f})
as
\beq
\dot\alpha+\frac{\dot f}{3g}=0.
\eeq
Taking the spatially projected derivative, one gets
\beq
\label{D_conserv_f}
D_a\dot\alpha- \frac{\dot\alpha}{\dot f}D_a\dot f+\frac{\dot\alpha}{g}
D_ag=0.
\eeq

We now wish to invert the time derivative and the spatial
gradient. In order to do so, it is  convenient to introduce the
Lie derivative along $u^a$, $\L$. Its action on a covector is given by
the expression
\beq
\L \chi_a \equiv u^c \nabla_c \chi_a + \chi_{c} \nabla_a u^c =
u^c
\partial_c \chi_a + \chi_{c} \partial_a u^c. \label{Lie_def}
\eeq
 For a scalar, $\L
f=\dot f$. The Lie derivation along $u^a$ and the spatial gradient
$D_a$ do not commute. Instead, one finds \cite{lv05a,lv05b}
\beq
\label{Ddot}
D_a\left(\dot f\right) =
\L \left(D_a f\right)- \dot u_a \dot f.
\eeq
Applying this identity both to $\alpha$ and $f$, one obtains
\beq
D_a\dot\alpha- \frac{\dot\alpha}{\dot f}D_a\dot f
= \L \left(D_a \alpha\right)- \frac{\dot\alpha}{\dot f}\L \left(D_a f\right).
\eeq
Substituting in (\ref{D_conserv_f}), one finds
\beq
\label{D_conserv_f2}
\L \left(D_a \alpha- \frac{\dot\alpha}{\dot f}D_a f\right)
 +\Tdot{\left(\frac{\dot\alpha}{\dot f}\right)} D_a f-3
\frac{\dot\alpha^2}{\dot f}
D_ag=0,
\eeq
where we have used Eq.~(\ref{conserv_f}) to rewrite  the last term.
Moreover, Eq.~(\ref{conserv_f}) implies
\beq
\Tdot{\left(\frac{\dot\alpha}{\dot f}\right)}=3 \dot{g}
\frac{\dot\alpha^2}{\dot f},
\eeq
which can be used to rewrite Eq.~(\ref{D_conserv_f2}) in the form
given in Eq.~(\ref{D_conserv_f3}).

For practical purposes, it is useful to note that, for the
covectors
\beq
\label{zeta_f}
\zeta_a^{(f)}=D_a \alpha- \frac{\dot\alpha}{\dot f}D_a f,
\qquad
\Gamma_a^{(g,f)}=D_a g- \frac{\dot g}{\dot f}D_af,
\eeq
which are defined as very particular linear combinations of {\it
spatially projected} gradients, one can replace the latter by {\it
ordinary} gradients and write
 \beq
 \label{partial}
\zeta_a^{(f)}=\partial_a \alpha- \frac{\dot\alpha}{\dot f}\partial_a f,
\qquad
\Gamma_a^{(g,f)}=\partial_a g- \frac{\dot g}{\dot f}\partial_af.
\eeq
This is a consequence of the identity
\beq
D_a \chi = \partial_a \chi + u_a \dot \chi,
\eeq
valid for any scalar quantity $\chi$.

\section{Nonlinear (non-)conservation equations}
\label{sec:nonconservation}

That one can construct conserved cosmological perturbations
associated with each quantity whose local evolution is determined
entirely by the local expansion of the universe was shown in
\cite{Lyth:2003im}. There, it was pointed out that such a
construction can be extended even in the nonlinear regime,
although an explicit expression of the nonlinear perturbation
variables was not given.

Here we explicitly construct these variables and their evolution
equations. In particular, we use the identity (\ref{D_conserv_f3})
to derive conservation -- or non-con\-ser\-va\-tion -- equations
for various covectors which represent nonlinear perturbations. In
this section, we discuss these equations for the nonlinear
generalization of the curvature perturbation on hypersurfaces of
uniform number density $n$, uniform energy density $\rho$, or
uniform entropy density $s$. The corresponding covectors are
obtained immediately by replacing the quantity $f$ in
Eq.~(\ref{D_conserv_f3}) with $n$, $\rho$, or $s$, respectively.
As in \cite{lv05a,lv05b}, the curvature perturbation is replaced
here by $\alpha$, the local number of e-folds of an observer
comoving with the fluid particles.

\subsection{Particle conservation}
The particle conservation equation,
\beq
\nabla_a (n\, u^a)=0,
\eeq
can be rewritten as
\beq
\dot n+\Theta n=0, \label{particle_cons}
\eeq
which is exactly of the form (\ref{conserv_f}) with
$f=g=n$.
Equation (\ref{D_conserv_f3})
tells us immediately that the particle conservation law
yields, for the covector
\beq
\zeta_a^{(n)}= D_a \alpha -{\dot\alpha\over \dot n}D_an,
\eeq
the equation
\beq
\L\zeta_a^{(n)}=0.
\eeq
This result has already been given in \cite{lv05b}.

This can be generalized to the case where the number of particles
is not conserved, in which case one can write
\beq
\nabla_a (n\, u^a)=-\Gamma n,
\eeq
where $\Gamma$ is the decay rate, which is not supposed to be a constant here.
Equivalently, one can write
\beq
\dot n+\Theta n+\Gamma n =0,
\eeq
which is still of the form (\ref{conserv_f}) with $f=n$ and
$g=n+(\Gamma n)/\Theta$. Taking the spatially projected gradients,
one thus finds
\beq
\L\zeta_a^{(n)}=
\frac{3 \dot \alpha^2}{\dot n} \left(
D_a \Gamma - \frac{\dot \Gamma}{\dot n} D_a n
- D_a \Theta + \frac{\dot \Theta}{\dot n} D_a n\right).
\eeq
The particle annihilation (or production) rate  $\Gamma n$ acts
here as a source for the evolution of the particle number density
perturbation. We will discuss more thoroughly the case of several
interacting fluids  below in Sec.~\ref{sec:interacting}.

\subsection{Energy conservation}
\label{sec:energy_cons}


In the general case, if one introduces  the quantity
\beq
\beta \equiv -\frac{\D}{\Theta}=
(\Theta\Pi+\pi^{ab} \sigma_{ab} + D_a q^a + 2 q^a \dot u_a )/{\Theta},
\eeq
then the energy conservation equation (\ref{energy_cons}),  becomes
\beq
\label{continuity} \dot\rho + \Theta (\rho + \p + \beta)=0,
\eeq
where $\beta$ acts as an extra pressure term, which  we will call
{\em dissipative} pressure.

One can apply the identity derived in the previous section with
$f=\rho$ and $g=\rho+\p+\beta$. This yields directly
\beq
\L\zeta_a= \frac{3 \dot \alpha^2}{ \dot \rho}\left(D_a \p
-\frac{\dot \p}{\dot \rho} D_a\rho  + D_a \beta - \frac{\dot
\beta}{\dot \rho} D_a\rho \right),
 \label{conserv1}
\eeq
in terms of the  covector  $\zeta_a$ defined in Eq.~(\ref{zeta2})
and introduced in \cite{lv05a,lv05b}.

Equation (\ref{conserv1}) is fully {\it nonperturbative} and
valid at {\it all scales}. It holds for any fluid, including
dissipative fluids with nonvanishing energy flow and anisotropic
stress. It can be rewritten as
\beq
\L \zeta_a= \frac{3 \dot \alpha^2}{ \dot \rho}
\left(\Gamma_{a}+\Sigma_{a}\right) , \label{conserv2}
\eeq
where, on the right hand side, one recognizes two covectors: the
nonlinear nonadiabatic pressure perturbation,
\beq
 \label{Gamma}
 \Gamma_a\equiv
D_a \p- {\dot \p\over \dot\rho}D_a\rho,
\eeq
which vanishes for purely adiabatic perturbations, i.e., when the
pressure $\p$ is solely a function of the energy density $\rho$,
and a term combining the gradients of the dissipative pressure and
of the energy density,
\beq
 \label{Sigma}
 \Sigma_a\equiv
D_a \beta- {\dot \beta\over \dot\rho}D_a\rho\, ,
\eeq
which we will call {\em dissipative} nonadiabatic pressure perturbation.
This vanishes for a purely perfect fluid.

Note that since the dissipative pressure $\beta$ depends on the
local expansion $\Theta$, the dissipative nonadiabatic pressure
perturbation $\Sigma_a$ depends implicitly on the local expansion.
In the appendix, Sec.~\ref{sec:alternative}, we discuss an
alternative formulation of the non-conservation equations which
leads to evolution equations where the source terms do not depend
on $\Theta$.

\subsection{Entropy (non-)conservation}

Defining
\beq
\tilde\beta \equiv -\frac{\D}{\Theta T}=\frac{\beta}{T},
\eeq
one recognizes in (\ref{s_dot}) an equation of the form (\ref{conserv_f})
with $f=s$ and $g=s+\tilde\beta$. One can thus write the evolution
equation
\beq
\L\zeta^{(s)}_a= \frac{3 \dot\alpha^2}{\dot s}
\left(D_a \tilde\beta - \frac{\dot{\tilde\beta}}{\dot s} D_a s \right),
\eeq
in terms of the  covector
\beq
 \zeta^{(s)}_a\equiv
D_a\alpha-\frac{\dot\alpha}{\dot s}D_a s,
\eeq
 which can be seen as the nonlinear generalization of the linear
curvature perturbation on constant entropy hypersurfaces.

\section{Interacting fluids}
\label{sec:interacting}

We now extend our  nonlinear formalism  to  a system
of interacting fluids.
Our treatment follows the approach of \cite{Malik:2002jb,Malik:2004tf},
in the context of the linear theory, and can thus be seen as
a nonlinear generalization of these works. In particular, we
extend to the non-linear case their study of  the coupled
evolution of curvature and {\em nonadiabatic} perturbations in a multifluid
system when energy transfer between the fluids is included.
\def\I{\alpha}
\def\J{\beta}
\def\It{{(\alpha)}}
\def\Jt{{(\beta)}}
\def\Q{{\cal Q}}
\def\S{{\cal S}}

We work in a common ``global'' frame defined by a unit
four-velocity $u^a$. The four-velocity can be conveniently chosen
depending on the physical problem (see \cite{Dunsby:1991xk} for a
discussion on this point). In this frame, the energy-momentum
tensor of each individual fluid can be expressed as
\beq
T^{\It}_{ab}= \rho^{\I}u_a u_b+ \p^{\I} h_{ab}+ q^{\It}_{a} u_{b}+
q^{\It}_b u_a +\pi^{\It}_{ab}.
\eeq
In the appendix, in Sec.~\ref{app:emt}, we give explicitly the
transformations from the global frame to each individual
$\I$-fluid frame.  Here, for simplicity, $ \p^{\I}$
denotes the total effective pressure for each fluid and we will
not include $\Pi^\I$ in the dissipative terms.

For each fluid, the (non-)conservation of the energy-momentum
tensor reads
\beq
\nabla_a T^\It{}^{ab}= Q^\It{}^b,
\eeq
where  $Q^\It{}^a$ is the energy-momentum transfer to the
$\I$-fluid. The conservation of the {\em total} energy-momentum
tensor implies the constraint
\beq
\label{sum_Q}
\sum_\I Q^\It{}^a = 0.
\eeq
Projecting along the four-velocity $u^a$ yields an energy
conservation equation,
\beq
\dot\rho^\I + \Theta (\rho^\I + \p^\I)= {\cal D}^\I + \Q^\I,
\label{energy_cons_I}
\eeq
with
\bea
\Q^\I &\equiv& - Q_b^\It u^b,  \\
{\cal D}^\I &\equiv& -\left( \pi^\It_{ab} \sigma^{ab} + D_a
q^\It{}^a + 2 q^\It{}^a \dot u_a \right), \label{Q_alpha}
\eea
where the dot and the spatial derivative $D_a$ are now defined
with respect to the four-velocity $u^a$ which a priori does not
coincide with any fluid frame.

If one then introduces  the  dissipative pressure of the $\alpha$-fluid,
\beq
\beta^\I \equiv -\frac{\D^\I}{\Theta}, \label{58}
\eeq
the energy conservation equation (\ref{energy_cons_I}),
becomes of the form (\ref{continuity}), with $\beta^\I -
\Q^\I/\Theta$ acting as an extra pressure term. This yields
\bea
\L\zeta^\It_a &=& \frac{3 \dot \alpha^2}{ \dot \rho^\I} \left( D_a
\p^\I - \frac{\dot \p^\I}{\dot \rho^\I} D_a \rho^\I + D_a \beta^\I
- \frac{\dot \beta^\I}{\dot \rho^\I} D_a \rho^\I
\right) \nonumber \\
&& - \frac{\dot \alpha}{ \dot \rho^\I}\left( D_a \Q^\I -
\frac{\dot \Q^\I }{\dot \rho^\I} D_a \rho^\I \right) +
\frac{\Q^\I}{3 \dot \rho^\I}\left( D_a \Theta - \frac{\dot
\Theta}{\dot \rho^\I} D_a \rho^\I\right) \label{zeta_multi1}.
\eea
This is the (non-)conservation equation for
\beq
\zeta^\It_a \equiv D_a \alpha -  \frac{\dot \alpha}{\dot \rho^\I}
D_a \rho^\I, \label{zeta_multi2}
\eeq
the nonlinear generalization of the  curvature
perturbation on uniform density hypersurfaces for the fluid $\I$.
This perturbation variable associated to the individual fluid $\I$
is conserved if the fluid is barotropic, $P^\I=P^\I(\rho^\I)$,
perfect, ${\cal D}^\I=0$, and decoupled from the other fluids,
${\cal Q}^\I=0$.

Note that $\zeta^\It_a$ is defined here with respect to the
four-velocity $u^a$ and not with respect to its own four-velocity
$u^{{\It}a}$, as it was the case in (\ref{zeta}). This does not
really affect the spatial gradients because they can be replaced,
as  before, by ordinary gradients.
 Nonetheless, the coefficient ${\dot \alpha}/{\dot \rho^\I}$
is different in the two cases, and one may have different
definitions of $\zeta^\It_a$ depending on $u^a$.
As discussed in the appendix Sec.~\ref{app:zeta}, in the
situations where the fluid relative velocities are small and the
geometry quasi-homogeneous, as in the cosmological context, then
the  various definitions of $\zeta^\I$ are equivalent in the {\it
linear} theory. Their spatial projections (i.e., perpendicular to
$u^a$)  are even equivalent up to {\em second} order.

To make the connection with the linear theory, in particular with
\cite{Malik:2002jb,Malik:2004tf}, one can rewrite
Eq.~(\ref{zeta_multi1}) as
\beq
\L\zeta^\It_a = \frac{3 \dot \alpha^2}{ \dot \rho^\I} \left(
\Gamma_a^\It + \Sigma_a^\It \right) - \frac{\dot \alpha}{ \dot
\rho^\I} \left( \Q_{a}^{(\I,{\rm intr})}   +\Q_{a}^{(\I,{\rm
rel})}   \right), \label{zeta_a_evol}
\eeq
where  we have identified several individual source terms: the
intrinsic nonadiabatic pressure perturbation,
\beq
\Gamma_a^\It \equiv D_a \p^\I - \frac{\dot \p^\I}{\dot \rho^\I}
D_a \rho^\I,
\eeq
the dissipative nonadiabatic pressure perturbation,
\beq
\Sigma_a^\It \equiv D_a \beta^\I - \frac{\dot \beta^\I}{\dot
\rho^\I} D_a \rho^\I,
\eeq
the intrinsic nonadiabatic energy transfer,
\beq
\Q_{a}^{(\I,{\rm intr})} \equiv D_a \Q^\I - \frac{\dot \Q^\I
}{\dot \rho^\I} D_a \rho^\I,
\eeq
and the relative nonadiabatic energy transfer,
\beq
\Q_{a}^{(\I,{\rm rel})} \equiv -  \frac{\Q^\I}{\Theta}\left(D_a
\Theta - \frac{\dot \Theta}{\dot \rho^\I} D_a \rho^\I \right).
\label{Q_rel}
\eeq

It is convenient, at this stage, to introduce the total
energy density and pressure,
\beq
\rho=\sum_\I \rho^\I, \qquad \p=\sum_\I \p^\I,
\eeq
 which are defined with respect to our unspecified global frame
$u^a$.
By summing the individual energy conservation equations (\ref{energy_cons_I}), one
gets an equation as in (\ref{energy_cons}) with
\beq
{\cal D}= \sum_\I {\cal D}^\I.
\eeq
The sum of the $\Q^\I$ vanishes as a consequence of the constraint (\ref{sum_Q}).

The covector $\zeta_a$ corresponding to $\rho$ can be expressed as
the  weighted sum of the individual $\zeta^\It_a$,
\beq
\zeta_a = \sum_\I\frac{\dot \rho^\I}{\dot \rho} \zeta^\It_a,
\eeq
and its evolution equation is given by Eq.~(\ref{conserv1}).
 Since now the global fluid is made of individual fluids, it
is useful to  split the global nonadiabatic pressure $\Gamma_a$
into two terms,
\beq
\Gamma_a = \Gamma^{({\rm intr})}_a + \Gamma^{({\rm rel})}_a
\label{Gamma_sum},
\eeq
 where the {\it intrinsic} nonadiabatic pressure
perturbation is the sum of the individual nonadiabatic pressure perturbations,
\beq
\Gamma^{({\rm intr})}_a \equiv \sum_\I \Gamma^{\It}_a.
\eeq
The {\it relative} adiabatic pressure perturbation can be written in
the form
\beq
\Gamma^{({\rm rel})}_a \equiv -\frac{1}{2 \Theta \dot \rho}
\sum_{\I,\J} \dot \rho^\I \dot \rho^\J \left(\frac{\dot
\p^\I}{\dot \rho^\I} - \frac{\dot \p^\J}{\dot \rho^\J}   \right)
\S_a^{(\I \J)}, \label{Gamma_rel}
\eeq
where  the relative entropy perturbations between the $\alpha$-
and $\beta$-fluids $\S_a^{(\I \J)}$ is defined as
\cite{Malik:2002jb}
\beq
\S_a^{(\I \J)} \equiv 3 \left( \zeta_a^\It - \zeta_a^\Jt \right) =
- 3 \dot \alpha \left( \frac{D_a \rho^\I}{\dot \rho^\I} -
\frac{D_a \rho^\J}{\dot \rho^\J} \right). \label{S_def}
\eeq
In order to derive Eq.~(\ref{Gamma_sum}) from the definition
(\ref{Gamma_rel}), it is convenient to use
\beq
\zeta_a^\It=\zeta_a + \frac{1}{3} \sum_\J \frac{\dot \rho^\J}{\dot
\rho} \S^{(\I \J)}. \label{zeta_a_zeta}
\eeq
Note that one can replace the spatial gradients in
Eq.~(\ref{S_def}) with partial derivatives.

A similar  decomposition applies to the dissipative nonadiabatic
pressure perturbation $\Sigma_a$, which can be written as
\beq
\Sigma_a = \Sigma^{({\rm intr})}_a + \Sigma^{({\rm rel})}_a,
\eeq
with the intrinsic part,
\beq
\Sigma^{({\rm intr})}_a \equiv \sum_\I \Sigma^{\It}_a,
\eeq
and the relative part,
\beq
\Sigma^{({\rm rel})}_a \equiv - \frac{1}{2 \Theta \dot \rho}
\sum_{\I,\J} \dot \rho^\I \dot \rho^\J \left(\frac{\dot
\beta^\I}{\dot \rho^\I} - \frac{\dot \beta^\J}{\dot \rho^\J}   \right)
\S_a^{(\I \J)}. \label{SigmaS}
\eeq
Note that one could also work directly with ${\cal D}^\I$
instead of $\beta^\I$, these two quantities being related by 
Eq.~(\ref{58}). 
As a consequence, one could separate each
corresponding covector $\Sigma_a^\It$ into an intrinsic and a relative part, in
analogy with our treatment of $\Q^\It_a$.

Let us now rewrite the relative nonadiabatic energy transfer of
the $\I$-fluid of Eq.~(\ref{Q_rel}) as
\beq
\Q^{(\I,{\rm rel})}_a = -\frac{\Q^\I}{\Theta} \left( D_a \Theta -
\frac{\dot \Theta}{\dot \rho} D_a \rho \right) - \frac{ \Q^\I \dot
\Theta}{\Theta^2} \sum_\J \frac{\dot \rho^\J}{\dot \rho} \S_a^{(\I
\J)},
\eeq
where we have employed Eq.~(\ref{zeta_a_zeta}) for the last term.
 By taking the difference
between the evolution equations (\ref{zeta_a_evol}) for two
fluids, we finally obtain an evolution equation for the relative
entropy perturbation,
\bea
\L \S_a^{(\I \J)} = \Theta^2 \left( \frac{\Gamma_a^{\It}
+ \Sigma_a^{\It}}{\dot \rho^\I}
- \frac{\Gamma_a^{\Jt} + \Sigma_a^{\Jt}}{\dot \rho^\J} \right) -
\Theta \left( \frac{\Q^{(\I, {\rm intr})}_a}{\dot
\rho^\I}- \frac{\Q^{(\J, {\rm intr})}_a}{\dot
\rho^\J} \right)
\quad \nonumber \\
+ \frac{\dot \Theta}{\Theta \dot \rho} \sum_\gamma \dot
\rho^\gamma \left( \frac{\Q^\I}{\dot \rho^\I}\S_a^{(\I \gamma)} -
\frac{\Q^\J}{\dot \rho^\J}\S_a^{(\J \gamma)} \right) +
\left(\frac{\Q^\I}{\dot \rho^\I} -\frac{\Q^\J}{\dot \rho^\J}
\right) \left( D_a \Theta - \frac{\dot \Theta}{\dot \rho} D_a \rho
\right). \nonumber
\\ \label{S_evolu}
\eea
The above equation represents the nonlinear generalization of
the evolution equation for the relative entropy perturbation, established in
\cite{Malik:2004tf} in the context of the linear theory.

Note that, while the total curvature perturbation $\zeta_a$ is
sourced by the relative entropy perturbations between the fluids
$\S_a^{(\I \J)}$, through Eqs.~(\ref{Gamma_rel}) and
(\ref{SigmaS}), the $\zeta_a^\It$'s do not appear in  the
evolution equation for the relative entropy perturbation
$\S_a^{(\I \J)}$. However,  $\S_a^{(\I \J)}$ is sourced by the
last term of Eq.~(\ref{S_evolu}), which depends on $\Theta$, i.e.,
on the local expansion. In the linear theory, this term vanishes
on large scales, i.e., on scales larger than the Hubble radius.

\section{Conclusion}
\label{sec:conclusion}

In the present work, we have extended our covariant formulation for
nonlinear perturbations
to the case of dissipative relativistic  fluids, allowing for
interactions.
This extension could be used to tackle more sophisticated physical situations.
Cosmology is such an example since the matter content of the universe is made
of  several fluids which can interact.

 In our approach, the
important quantities are the covectors
$\zeta_a^{(f)}$, defined as linear
combinations of the spatial
gradients of  the number of e-folds $\alpha$, and of some
 scalar fluid quantities $f$,
for one or several fluids.
These covectors are fully nonlinear and
generalize the curvature perturbations on
hypersurfaces of constant $f$
defined in the context of linear cosmological perturbation theory.

The non-conservation equation for $\zeta_a$ associated with the
energy density, that we obtain when the fluid is unperfect, can be
derived and written in a form analogous to that found in our
initial formalism with a single perfect fluid, provided that the
pressure is modified such as to include dissipative effects.

For several interacting fluids, we can also define fully nonlinear
quantities that generalize analogous quantities introduced in the
linear theory. Remarkably, as in our initial formalism with a
single perfect fluid, the evolution equations that govern these
quantities ``mimic'' those found in the linear theory, with the
advantage that they are covariant, i.e., independent of any
coordinate system. Thus, it is straightforward to linearize them
or expand them at second, or higher,  order in the perturbations.

Moreover, our fully nonlinear approach allows to identify which
property is specific to linear  (or second) order and
which property will remain valid at all orders.  For example,
in the multifluid case, there are a priori several  nonlinear
generalizations of the
curvature perturbation, which depend on the choice of the reference
frame.  In the cosmological context, they all reduce to the same
quantity at linear order.

\appendix

\section{Alternative formulation of the non-con\-ser\-va\-tion equations}
\label{sec:alternative}
In order to apply our general identity (\ref{D_conserv_f3})
to the case of a dissipative
fluid, we have introduced the dissipative pressure $\beta$ which depends
on the local expansion $\Theta$. In doing this, we have obtained
equations like (\ref{conserv1}) where the source covectors,
on the right hand side,
depend implicitly on the local expansion.
As we show below, there exists  an alternative formulation, which leads
to evolution equations where the source terms do not depend on $\Theta$.

Instead of writing the non-conservation equations
in the form (\ref{conserv_f}), an alternative possibility is to
start from an equation of the form
\beq
\label{conserv_fB}
\dot f+\Theta g=h.
\eeq

Taking the spatially projected derivative of the above equation
and replacing $\Theta$ by $3\dot\alpha$, one gets
\beq
\label{D_conserv_fB}
D_a\dot f+3 gD_a\dot\alpha +3\dot\alpha D_ag=
D_ah.
\eeq

 For the first two terms one can invert the gradient and time
derivative by using the identity (\ref{Ddot}). One obtains
\beq
\L \left(D_a f\right)+3g \L \left(D_a \alpha\right) = D_a h - 3
\dot\alpha D_a g + h\dot u_a,
\eeq
where we have used Eq.~(\ref{conserv_fB}) to simplify the terms
proportional to $\dot u_a$. Using again Eq.~(\ref{conserv_fB}) to
replace $3 \dot \alpha$ on the right hand side, this relation can
be rewritten after some straightforward manipulations in the form
\beq
\L \left(D_a \alpha+ \frac{1}{3g}D_a f\right)=-\frac{1}{3g^2}\left(\dot g D_af-\dot fD_a g
+h D_a g-g D_a h\right)+\frac{h}{3g}\dot u_a.
\label{D_conserv_f3B}
\eeq

We have thus derived an evolution equation for the covector
\beq
{\hat \zeta}_a^{(f)}\equiv D_a \alpha+ \frac{1}{3g}D_a f,
\eeq
which is a linear combination of the spatial gradients of $\alpha$ and $f$,
but which differs from the covector $\zeta_a^{(f)}$,
defined in (\ref{zeta_f}), when $h$ in
(\ref{conserv_fB}) does not vanish.
 These two quantities have also  different geometrical
interpretations. In particular, unless $h=0$,
${\hat \zeta}_a^{(f)}$ {\em does not} generalize the curvature perturbation
on the hypersurface of uniform $f$, as $\zeta_a^{(f)}$ does.
However, one advantage of this alternative formulation is that the
number of e-folds $\alpha$, or its  proper time derivative, the
expansion $\Theta$, appears only on the left hand side of the
non-conservation equations. On the right hand side, one only finds
quantities characterizing the fluid, although the acceleration now
appears.

This new identity can be applied  to the case where the number of particles
is not conserved, in which case one can write
\beq
\dot n+\Theta n=\Xi,
\eeq
where $\Xi$ is the production rate of particles.
Eq.~(\ref{D_conserv_f3B}) with $f=g=n$ and $h=\Xi$ then yields directly
\beq
\L{\hat\zeta}_a^{(n)}= -\frac{\Xi}{3n} \left(
\frac{D_a n}{n}-\frac{D_a\Xi}{\Xi} - \dot u_a\right).
\eeq

The new identity also applies to the non-conservation of energy in
the dissipative case, with $f=\rho$, $g=\rho+\p$ and $h={\cal D}$.
This yields directly, for the covector
\beq
{\hat\zeta}_a^{(\rho)}\equiv D_a \alpha+ \frac{1}{3(\rho+\p)}D_a
\rho,
\eeq
the evolution equation
\bea
\L {\hat\zeta}_a^{(\rho)}&=&-\frac{1}{3(\rho+\p)^2}\left[\dot \p
D_a\rho-\dot\rho D_a \p +{\cal D} D_a (\rho+\p) \right.\nonumber
\\&& \left. -(\rho+\p) D_a {\cal D}\right]+\frac{{\cal
D}}{3(\rho+\p)}\dot u_a.
 \label{conserv1B}
\eea
The first two terms on the right hand side are proportional to the covector
$\Gamma_a$ defined in (\ref{Gamma}).

\section{Multifluid system}

\subsection{Frame dependency of the energy-momentum tensor}
\label{app:emt}

In a multifluid system,
the energy-momentum tensor of each fluid $\I$ can be written in
the form
\beq
T^{\It}_{ab}= \tilde \rho^{\I}   u^{\It}_a u^{\It}_b+ \tilde
\p^{\I} h^\It_{ab}+ \tilde q^{\It}_{a} u^{\It}_{b}+ \tilde
q^{\It}_b u^{ \It}_a + \tilde \pi^{\It}_{ab}, \label{rest}
\eeq
where $u^{\It}{}^a$ is the unit four-velocity of the fluid $\I$ --
e.g., the rest frame of the particles of the fluid $\I$ -- and $
h^\It_{ab}$ is the projector on the hypersurface orthogonal to
$u^{\It}{}^a$,
\beq
h^\It_{ab} \equiv g_{ab} + u_a^\It u_b^\It. \label{h_alpha}
\eeq
Quantities with the tilde are written in the $\I$-fluid frame.

Each fluid can have a different frame $u^{\It}{}^a$. It is
preferable, however, to work in an arbitrary global frame defined
by a common unit four-velocity $u^a$. This is related to the
four-velocity of each individual fluid by the relation
\beq
u^{\It}_a = \gamma_{\I} \left(u_a + v^{\It}_a \right), \quad \quad
u^a v_a^{\It}=0,
\eeq
with the relativistic factor $\gamma_\I = (1-g^{ab}v_a^\It
v_b^{\It})^{-1/2}$ and $v^{\It}{}^a$ a peculiar velocity. The
energy momentum tensor of each individual fluid can be rewritten
in the global frame as
\beq
T^{\It}_{ab}= \rho^{\I}  u_a u_b+ \p^{\I} h_{ab}+ q^{\It}_{a}
u_{b}+  q^{\It}_b u_a +\pi^{\It}_{ab}.
\eeq
The quantities in the (untilted) global frame can be expressed in
terms of the tilted quantities and of the peculiar velocities. The
corresponding expressions  are given by \cite{Roy} and read
\bea
\rho^\I &=& \gamma_\I^2 \tilde \rho^\I + \left(\gamma_\I^2-1
\right) \tilde \p^\I + 2 \gamma_\I \tilde q^\It_a v^\It{}^a +
\tilde \pi_{ab}^\It v^\It{}^a
v^\It{}^b, \\
\p^\I&=&  \frac{1}{3} \left(\gamma_\I^2-1 \right) \left( \tilde
\rho^\I+ \tilde \p^\I \right) + \tilde \p^\I + \frac{2}{3}
\gamma_\I \tilde q^\It_a v^\It{}^a + \frac{1}{3} \tilde
\pi_{ab}^\It v^\It{}^a
v^\It{}^b, \\
q^\It_a &= &  \gamma_\I^2 \left( \tilde \rho^\I+ \tilde \p^\I
\right) v_a^\It + \gamma_\I \tilde q^\It_a +\gamma_\I \tilde
q^\It_b v^\It{}^b \left( v_a^\It - u_a \right) \nonumber \\ &&+
\tilde \pi^\It_{ab} v^\It{}^b  - \tilde \pi^\It_{bc} v^\It{}^b
v^\It{}^c u_a, \\
\pi^\It_{ab}  &=& \gamma_\I^2 \left( \tilde \rho^\I+ \tilde \p^\I
\right) \left(v_a^\It v_b^\It - \frac{1}{3}h_{ab} v_c^\It
v^c{}^\It \right) \nonumber \\&&+ \gamma_\I \left( \tilde q^\It_a
v^\It_b +\tilde q^\It_b v^\It_a
- \frac{2}{3}h_{ab} \tilde q^\It_c v^\It{}^c \right) \nonumber \\
&&- \gamma_\I \tilde q^\It_c v^\It{}^c \left(u_a v^\It_b + u_b
v^\It_a
 \right) +\tilde \pi_{ab} -\left(u_a \tilde \pi_{bc}+ u_b \tilde \pi_{ac}\right)v^c{}^\It
 \nonumber \\&&+ \tilde \pi^\It_{cd}  v^\It{}^c v^\It{}^d u_a u_b-
 \frac{1}{3} h_{ab} \tilde \pi^\It_{cd}  v^\It{}^c v^\It{}^d.
\eea
These equations allow to reexpress in the global frame defined by
$u^a$, the energy-momentum tensor of each fluid $\I$ written in
their rest frame $u^{\It}{}^a$.

Note that at linear order in the perturbations, if the fluid
relative velocities are small and the geometry quasi-homogeneous,
as in the cosmological context, then  the energy density, pressure
and anisotropic stress of the fluids do not depend on the frame.
Only the energy flow depends on the peculiar velocity $v^\It{}^a$.

\subsection{Frame dependency of $\zeta_a^\It$}
\label{app:zeta}

The covector $\zeta_a^\It$ defined in Eq.~(\ref{zeta_multi2})
for a fluid $\alpha$
depends on the choice of the global four-velocity $u^a$. However, for
each individual fluid, one can define a nonlinear generalization
of the curvature perturbation on uniform density hypersurfaces,
only with respect to its own four-velocity $u^{{\It}a}$. This is
defined as
\beq
\tilde \zeta_a^\It \equiv D^\It_a \alpha - \frac{u^\It{}^b\nabla_b
\alpha}{u^\It{}^b\nabla_b \rho} D^\It_a \rho, \label{tilde_zeta_a}
\eeq
where
\beq
 D^\It_a \chi \equiv h^\It_{a}{}^{b} \nabla_b \chi,
\label{tilde_D}
\eeq
and $ h^\It_{ab}$ is  defined in Eq.~(\ref{h_alpha}). Note that one
can replace the spatial gradients with standard partial
derivatives in Eq.~(\ref{tilde_zeta_a}).

Let us now assume the situation where the relative velocities between the
fluids are small and let us consider a common reference velocity $u^a$
so that the differences
\beq
\epsilon^\It{}^a = u^\It{}^a - u^a
\eeq
are small. One can then treat the $\epsilon^\It{}^a$ as
perturbations and the expansion of the coefficient in
Eq.~(\ref{tilde_zeta_a}) yields, at first order in
$\epsilon^\It{}^a$,
\beq
\frac{u^\It{}^a\nabla_a \alpha}{u^\It{}^a\nabla_a \rho}
\simeq \frac{\dot \alpha}{\dot \rho} +
\frac{1}{\dot \rho} \epsilon^\It{}^a \zeta_a \, .
\eeq
Inserting this result into the definition (\ref{tilde_zeta_a}) of
$\tilde \zeta_a^\It$, one gets
\beq
\tilde \zeta_a^\It = \partial_a \alpha - \frac{u^\It{}^b\nabla_b
\alpha}{u^\It{}^b\nabla_b \rho} \partial_a \rho \simeq \zeta_a^\It
+ u_a \epsilon^\It{}^b \zeta_b^\It.
\eeq
Consequently, the projections of the two covectors $\tilde
\zeta_a^\It$ and $\zeta_a^\It$ perpendicular to $u^a$ are
equivalent up to  first order in $\epsilon^\It{}^a$.

One can go one step further when the spacetime is approximately
homogeneous, as in cosmology, because the spatial gradients can
then be considered as perturbations as well.  Then, the difference
between $\tilde \zeta_a^\It$ and $\zeta_a^\It$ is of second order.
This implies that spatial parts of the two covectors, i.e., their
projections perpendicular to $u^a$, are equivalent {\em up to}
second order.


\begin{thebibliography}{99}




\bibitem{lv05a}
  D.~Langlois and F.~Vernizzi,
  Phys.\ Rev.\ Lett.\  {\bf 95}, 091303 (2005)
  [arXiv:astro-ph/0503416].

\bibitem{lv05b}
  D.~Langlois and F.~Vernizzi,
  Phys.\ Rev.\ D {\bf 72}, 103501 (2005)
  [arXiv:astro-ph/0509078].

\bibitem{Hawking:1966qi}
  S.~W.~Hawking,
  Astrophys.\ J.\  {\bf 145}, 544 (1966).

\bibitem{ellis}
G.~F.~R.~Ellis, Relativistic Cosmology, in {\em General Relativity
and Cosmology}, proceedings of the XLVII Enrico Fermi Summer
School, edited by R.~K.~Sachs (Academic, New York, 1971).

\bibitem{Ellis:1989jt}
G.~F.~R.~Ellis and M.~Bruni,
Phys.\ Rev.\ D {\bf 40}, 1804 (1989).

\bibitem{Hwang:1990am}
  J.~c.~Hwang and E.~T.~Vishniac,
  Astrophys.\ J.\  {\bf 353}, 1 (1990).

\bibitem{Dunsby:1991vv}
  P.~K.~S.~Dunsby,
  Class.\ Quant.\ Grav.\  {\bf 8}, 1785 (1991).


\bibitem{Dunsby:1991xk}
  P.~K.~S.~Dunsby, M.~Bruni and G.~F.~R.~Ellis,
  Astrophys.\ J.\  {\bf 395}, 54 (1992).


\bibitem{Challinor}
  A.~Challinor and A.~Lasenby,
  Phys.\ Rev.\ D {\bf 58}, 023001 (1998)
  [arXiv:astro-ph/9804150].


\bibitem{Malik:2002jb}
  K.~A.~Malik, D.~Wands and C.~Ungarelli,
  Phys.\ Rev.\ D {\bf 67}, 063516 (2003)
  [arXiv:astro-ph/0211602].


\bibitem{Malik:2004tf}
  K.~A.~Malik and D.~Wands,
  JCAP {\bf 0502}, 007 (2005)
  [arXiv:astro-ph/0411703].


\bibitem{maartens}
  R.~Maartens,
  arXiv:astro-ph/9609119.

\bibitem{Israel:1979wp}
  W.~Israel and J.~M.~Stewart,
  Annals Phys.\  {\bf 118}, 341 (1979).

\bibitem{Lyth:2003im}
  D.~H.~Lyth and D.~Wands,
  Phys.\ Rev.\ D {\bf 68}, 103515 (2003)
  [arXiv:astro-ph/0306498].

\bibitem{Lyth:2004gb}
  D.~H.~Lyth, K.~A.~Malik and M.~Sasaki,
  JCAP {\bf 0505}, 004 (2005)
  [arXiv:astro-ph/0411220].

\bibitem{Rigopoulos:2003ak}
G.~I.~Rigopoulos and E.~P.~S.~Shellard,
Phys.\ Rev.\ D {\bf 68}, 123518  (2003) [arXiv:astro-ph/0306620].

\bibitem{Roy}
  R.~Maartens,
  Phys.\ Rev.\ D {\bf 58}, 124006 (1998)
  [arXiv:astro-ph/9808235].
















\end{thebibliography}
\end{document}